\documentclass[usenatbib,useAMS]{mn2e}
\usepackage{times}
\usepackage{epsfig}

\renewcommand{\vec}[1]{\bmath{#1}}

\newcommand{\fss}{\scriptscriptstyle}

\begin{document}
\title{Lens binarity vs limb darkening in close-impact galactic microlensing events}
\author[M. Dominik]{M. Dominik \\
University of St Andrews, School of Physics \& Astronomy, North Haugh,
St Andrews, KY16 9SS}

\maketitle

\begin{abstract}
Although point caustics harbour a larger potential for measuring the
brightness profile of stars during the course of a microlensing event than
(line-shaped) fold caustics, the effect of lens binarity significantly 
limits the achievable accuracy. Therefore, corresponding close-impact
events make a less favourable case for limb-darkening measurements than
those events that involve fold-caustic passages, from which precision
measurements can easily and routinely be obtained. Examples involving
later Bulge giants indicate that a $\sim 10\,\%$ misestimate
on the limb-darkening coefficient can result with the assumption of a single-lens model that
looks acceptable, unless the precision of the photometric measurements is pushed below
the $1\,\%$-level even for these favourable targets. In contrast, measurement uncertainties
on the proper motion between lens and source are dominated by the assessment of the 
angular radius of the source star and remain practically unaffected by lens binarity.
Rather than judging the goodness-of-fit by means of
a $\chi^2$ test only, run tests provide useful additional information that
can lead to the rejection of models and the detection of lens binarity in
close-impact microlensing events. 
\end{abstract}


\section{Introduction}

In order to resolve the surface of the observed star during a microlensing event, the
magnification pattern created by the lens needs to supply a large magnification gradient.
Two such configurations meeting this requirement have been discussed extensively 
in the literature:
a point-caustic at the
angular position of a single point-like lens 
\citep{WM94,NemWick94,Gould94:extsrc,BC95,BC96,Witt95,GW96,GG:detld,Heyetal2000,Hey2003} 
and a line-shaped fold caustic produced by a binary lens 
\citep{SchneiWei:QSO,SchneiWag, GG:detld,Rhie:ld,Do:SecondLD,Do:Fold,Do:FoldLD}. 
It has been pointed out by \citet{GG:detld} that fold-caustic events are more common and
their observation is easier to plan, whereas close-impact events where the source transits
a point caustic can provide more information. However, I will argue that 
this apparent gain of
information can usually not be realized
due to potential lens binarity. 
In contrast to fold caustics which form a generically stable singularity, point
caustics are not stable and do not exist in reality.
Instead, there is always a small
diamond-shaped caustic containing four cusps.

In this paper, the influence of lens binarity on
the measurement of stellar limb-darkening coefficients and proper motion is investigated 
and the arising limitations of the power of close-impact events where the source passes over
a single closed caustic are discussed.
Sect.~\ref{sec:cie} discusses the basics of close-impact microlensing events with the
effect of source size, the potential of measuring stellar proper motion and limb darkening, and
the effect of lens binarity. Sect.~\ref{sec:binvsld} shows the influence of lens binarity on
the extraction of information from such events. 
First, the effect of lens binarity on the 
light curves is demonstrated by means of two illustrative examples involving
K and M Bulge giants. Subsequently, a simulation of data corresponding to these 
configurations is used to investigate the potential misestimates of parameters if lens binarity
is neglected.
Sect.~\ref{sec:conclusions} presents the final conclusions and a summary of the results.

\section{Close-impact microlensing events}
\label{sec:cie}
\subsection{Size of source star}
As pointed out by~\citet{Pac86}, a point-like source star at a distance $D_\rmn{S}$ from the
observer exhibits a magnification due to the gravitational field of a lens star with mass $M$ at
$D_\rmn{L}$ by a factor
\begin{equation}
A(u) = \frac{u^2 + 2}{u\,\sqrt{u^2+4}}\,,
\label{eq:ptmag}
\end{equation}
where source and lens are separated by the angle $u\,\theta_\rmn{E}$ and 
\begin{equation}
\theta_\rmn{E} = \sqrt{\frac{4GM}{c^2}\,\frac{D_\rmn{S} - D_\rmn{L}}{D_\rmn{S}\,D_\rmn{L}}}
\end{equation}
denotes the angular Einstein radius.

The proper motion $\mu$ of the source relative to the lens constitutes
a microlensing event with the time-scale $t_\rmn{E} = \theta_\rmn{E}/\mu$, for which the
lens-source separation becomes
\begin{equation}
u(t) = \sqrt{u_0^2 + [p(t)]^2}\,,
\end{equation}
where $p(t) = (t-t_0)/t_\rmn{E}$, so that $u_0$ is the closest approach which occurs at time $t_0$.

It was discussed by \citet{WM94} as well as by \citet{NemWick94}
that the finite extent of the source star could cause
observable deviations from the magnification factor $A(u)$ as given by Eq.~(\ref{eq:ptmag}).
\citet{Gould94:extsrc} has argued that for a source star with radius $R_\star = \rho_\star\,
\theta_\rmn{E}$ and radial brightness profile $I(\rho) = \overline{I}\,\xi(\rho)$, where
$\rho$ is the fractional radius and $\overline{I}$ is the average brightness,
the finite-source magnification can be approximated as
\begin{equation}
A^\rmn{ext}(u, \rho_\star; \xi) = A(u)\,B(u/\rho_\star; \xi)\,,
\end{equation}
with 
\begin{equation}
B(z;\xi) = \frac{1}{\pi}\,
\int_{0}^{2\pi} \int_0^1 \xi(\rho)\,\frac{z}{\sqrt{\rho^2+2 \rho z
\cos\varphi + z^2}}\,\rho\,\mathrm{d}\rho\,\mathrm{d}\varphi\,.
\end{equation}
In particular, $B(z) \simeq 1$ for $z \gg 1$, so that for large angular separations (compared to
$\rho_\star$), the finite source effects becomes negligible, whereas $B(z) \simeq \beta z$ for 
$z \ll 1$ implies strong effects for small angular separations.
Therefore, microlensing events with small impact parameters $u_0 \ll \rho$ are the most likely
to show prominent effects of finite source size. Stellar spectra can be used to derive
radius $R_\star$ and distance $D_\rmn{S}$ of the source star, yielding its angular radius
$\theta_\star = R_\star/D_\rmn{S}$. On the other hand, microlensing observations yield 
$\rho_\star$ and therefore the time-scale $t_\star = \rho_\star\,t_\rmn{E}$, during which the
source moves by its own angular radius on the sky. Therefore, the observation of finite source
effects in microlensing events provides a measurement of the proper motion between lens and source
as
\begin{equation}
\mu = \frac{\theta_\star}{t_\star} = \frac{R_\star}{D_\rmn{S}\,\rho_\star\,t_\rmn{E}}\,.
\end{equation}

\subsection{Limb darkening}
As already indicated, the stellar surface is not uniformly bright, but a characteristic variation
with the distance from the center is observed, commonly known as limb darkening, which depends
on wavelength and therefore on the filter used for the observations.
A widely used model is the linear limb-darkening law \citep{Milne21} 
\begin{equation}
\xi(\rho) = 1+\Gamma\,\left(\frac{3}{2}\,\sqrt{1-\rho^2}-1\right)\,,
\end{equation}
with $0 \leq \Gamma \leq 1$,
which is linear in $\cos \vartheta = \sqrt{1-\rho^2}$, where $\vartheta$ is the emergent angle.
If the point-source magnification $A(u)$ shows a strong variation over the face of the source star,
dense and precise microlensing observations provide an opportunity for a measurement of the 
limb-darkening coefficient \citep{BC96,GW96}. 

The strongest magnification gradients occur in the vicinity of caustics. While a single lens
creates a point caustic at its angular position, binary lenses create finite caustic curves which
contain cusps. As the angular separation between the binary lens objects tends to zero, its
diamond-shaped caustic with four cusps degenerates into the point caustic of a single lens.
There are two different main scenarios 
with the potential of providing limb-darkening measurements: 
passages of the source star over the point caustic
of a single lens, so that the impact parameter falls below the angular source radius, i.e.\
$u_0 < \rho_\star$ 
\citep{BC96,GW96,GG:detld,Heyetal2000,Hey2003}, 
and passages of the source star over a (line-shaped) fold caustic
created by a binary lens 
\citep{SchneiWei:QSO,SchneiWag, GG:detld,Rhie:ld,Do:SecondLD,Do:FoldLD}. 
In addition, the source might pass directly over a 
cusp, as for the event MACHO 1997-BLG-28 \citep{PLANET:M28}, for which the first limb-darkening measurement
by microlensing has been obtained. As anticipated by \citet{GG:detld}, 
the vast majority of other limb-darkening measurements so far
has arisen from fold-caustic passages 
\citep{joint,PLANET:M41,PLANET:O23,PLANET:EB5}, whereas two measurements from single-lens
events have been reported so far \citep{MACHO:9530,Yooetal}.
The remaining limb-darkening measurement by microlensing, on the solar-like
star MOA 2002-BLG-33 \cite{Abe:KB0233} constituted a very special case, where
the source simultaneously enclosed several cusps over the course of its passage.

\subsection{Lens binarity}
Since the majority of stars resides in some form of binary or multiple systems \citep[e.g.][]{Abt},
it seems at first sight a bit surprising that more than 85~\% of the observed microlensing events
appear to be consistent with the assumption of a single point-like lens. 
An important clue to this puzzle is that the separation of binaries covers a broad range of
6--7 orders of magnitude roughly from contact to typical stellar distances. Lens binarity however 
usually does not provide strong deviations to the light curve if the angular separation is much
smaller or much larger than the angular Einstein radius $\theta_\rmn{E}$ \citep{MP91}. Therefore, many 
binary lenses simply escape our attention by failing to provide an observable signal
\citep{DiStef:Binaries}.
While several binary lens systems provide weak distortions, such as MACHO LMC-1
\citep{DoHi1,DoHi2} or MACHO 1999-BLG-47 \citep{PLANET:MB9947}, a characteristic signature is
provided if the source passes over a fold or even a cusp caustic, OGLE-7 \citep{OGLE7}
being the first such observed event.

Thus, although a lens star is never a completely isolated object, it can be approximated as such in
many cases. However, one needs to keep in mind that its caustic is a small diamond with four cusps
rather than a single point. While for a single lens, the computation of the light curve is
a rather straightforward process, where a semi-analytical expression by means of elliptical
integral exists for uniformly bright sources \citep{WM94}, it becomes quite demanding if a 
binary lens is considered due to the fact that no closed expression exists for the point-source
magnification of a binary lens, but in general a fifth-order polynomial needs to be solved
\citep{WM95:fifth,Asada:fifth}. For this paper, the algorithm of \citet{Do98:NumSrc} has been used
which is based on the contour plot technique by \citet{SK87} and the application of Green's
theorem.

Compared to a single lens, a binary lens involves three additional parameters, which can be
chosen as the mass ratio $q$, the angle $\alpha$, and the separation parameter $d$.
Here, $q = M_2/M_1$ is the ratio between the masses of secondary and the primary 
component which are separated on the sky by the angle $d\,\theta_\rmn{E}$, while 
$\alpha$ is measured from the vector pointing from the 
secondary to the primary towards the source trajectory which therefore reads
\begin{equation}
\vec u (t) = u_0 \left(\begin{array}{c}-\sin \alpha \\ \cos\alpha \end{array}\right) +
p(t) \left(\begin{array}{c}\cos \alpha \\ \sin\alpha \end{array}\right)\,.
\label{eq:trajectory}
\end{equation}

\section{Binarity vs limb darkening}
\label{sec:binvsld}

\subsection{Illustrative binary-lens configurations}
\label{subsec:examples}

\begin{table}
\caption{Model parameters and indicative corresponding physical properties 
for the two discussed event configurations.}
\label{tab:truemodel}
\begin{tabular}{@{}lcc}
\hline
& configuration I & configuration II \\
\hline
$t_\rmn{E}$~[d] & $55$ & $35$ \\
$t_0$~[d] & 0 & 0 \\
$u_0$ & $0.015$ & $0.05$ \\
$\rho_\star$ & $0.05$ & $0.075$ \\
$\alpha$ & $0^{\degr}$ & $0^{\degr}$ \\
$q$ & $1$ & $1$ \\
$I_\rmn{base}$ & $13.6$ & $12.3$ \\
$g$ & $0$ & $0$ \\
$\Gamma_I$ & $0.5$ & $0.5$ \\
\hline
$D_\rmn{S}$~[kpc] & $\sim\,8.5$ & $\sim\,8.5$ \\
$D_\rmn{L}$~[kpc] & $\sim\,6.5$ & $\sim\,6.5$ \\
$M$~[$M_{\sun}$] & $\sim\,0.35$ & $\sim\,0.7$ \\
$v$~[km\,s$^{-1}$] & $\sim\,55$ & $\sim\,140$ \\
$\mu$~[$\mu$as\,d$^{-1}$] & $\sim\,330$& $\sim\,460$\\
$\theta_\rmn{E}$~[$\mu$as] & $\sim\,6$ & $\sim\,15$\\
$r_\rmn{E}$~[AU] & $\sim\,2.0$ & $\sim\,2.9$ \\
$r_\rmn{E}'$~[$R_{\sun}$] & $\sim\,500$ & $\sim\,800$\\
$R_\star$~[$R_{\sun}$] & $\sim\,25$ & $\sim\,60$ \\
spectral type & K5III & M4III \\
$M_V$ & $-0.2$ & $+0.2$  \\
$V-I$ & $+2.0$ & $+3.4$  \\
$A_I$ & $1.25$ & $0.85$ \\
$m-M$ & $14.65$ & $14.65$ \\
\hline
\end{tabular}
\end{table}

\begin{figure*}
\includegraphics[width=84mm]{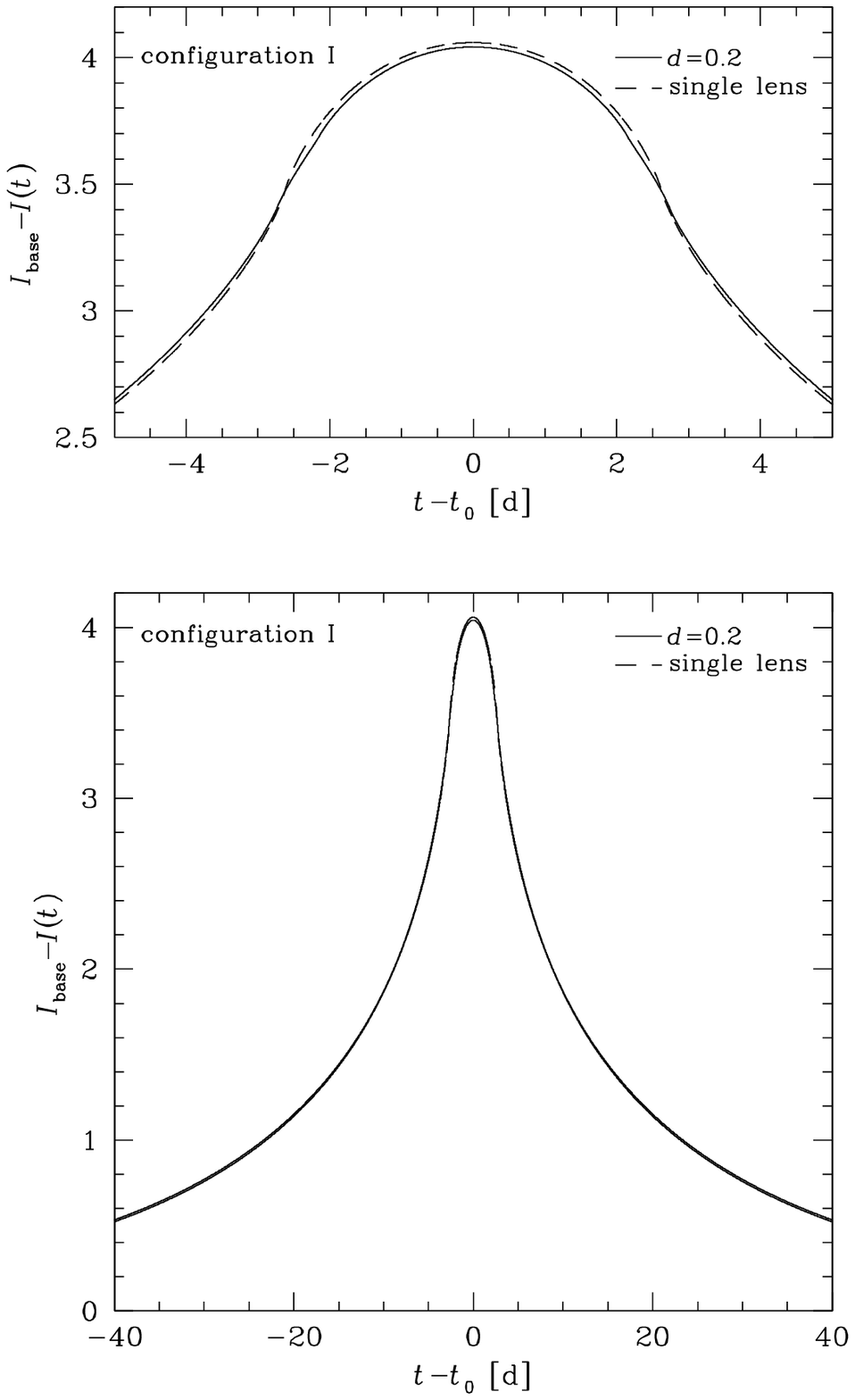}\hspace{8mm}
\includegraphics[width=84mm]{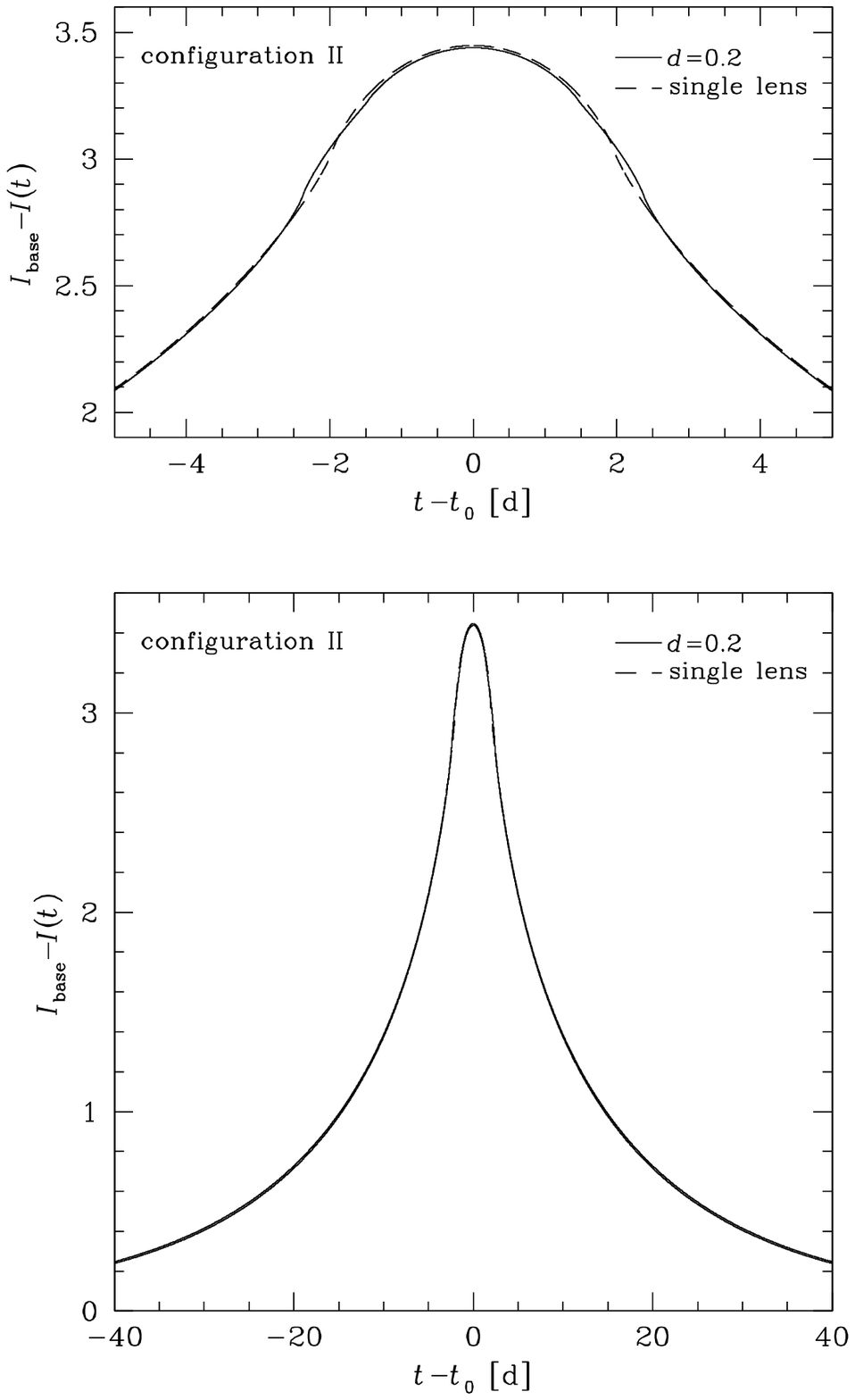}
\caption{Light curves corresponding to
the binary-lens configuration I (left) and configuration II (right) with parameters as listed in
Table~\ref{tab:truemodel} and a separation parameter $d=0.2$ (solid line) 
along with those for a single
lens model with otherwise identical parameters (dashed line). In the lower panels, which show the full
wing of the light curves, binary-lens and single-lens models are hardly distinguishable.}
\label{fig:curves}
\end{figure*}

In order to investigate the effect of lens binarity on the measurement of stellar limb darkening
and proper motion,
two different event configurations have been chosen as illustrative examples, where different
angular separations between the lens objects have been considered for both of them.
For the model referred to as 'configuration I', let us consider a K5 Bulge giant at $D_\rmn{S} \sim
8.5~\mbox{kpc}$, for which $M_V = -0.2$ and $V-I = +2.0$. One therefore 
obtains $I_0 = 12.35$, so that with an assumed extinction $A_I \sim 1.25$,
the observed source magnitude becomes $I \sim 13.6$.
Let us further assume a binary lens of total mass 
$M \sim 0.35~M_{\odot}$ at $D_\rmn{L} \sim 6.5~\mbox{kpc}$, so that the angular Einstein
radius becomes $\theta_\rmn{E} \sim\,320~\mu\mbox{as}$, the Einstein radius
becomes $r_\rmn{E} = D_\rmn{L}\,\theta_\rmn{E} \sim 2.0~\mbox{AU}$ and its projection to the
source distance becomes
$r_\rmn{E}' = (D_\rmn{S}/D_\rmn{L})\,r_\rmn{E} \sim 500~R_{\odot}$.
With $R_\star \sim 25~R_{\odot}$ for a K5 giant, 
a source size parameter $\rho_\star = 0.05$ is therefore adopted. In order to make an optimal
case for observing, a rather low proper motion $\mu \sim\,6~\mbox{km}\,\mbox{s}^{-1}$,
corresponding to a lens velocity of $v \sim 55~\mbox{km}\,\mbox{s}^{-1}$ relative
to the source has been chosen, which yields an event time-scale $t_\rmn{E} = 55~\mbox{d}$.
The impact parameter is chosen as $u_0 = 0.015$, corresponding to a peak magnification of
a point source of $A_0 \sim 70$.

Configuration II has been chosen to be similar to the parameters of the observed event
MACHO 1995-BLG-30 \citep{MACHO:9530}, for which the source is an even larger star, namely an
M4 giant. According to the obtained model parameters,
let us adopt $u_0 = 0.05$, $\rho_\star = 0.075$, and the event time-scale
$t_\rmn{E} = 35~\mbox{d}$. For $D_\rmn{S} \sim\,8.5~\mbox{kpc}$ and $D_\rmn{L} \sim\,6.5~\mbox{kpc}$,
the appropriate stellar radius of $R_\star \sim\,60~R_{\sun}$ is obtained for $M \sim\,0.7~M_{\sun}$,
so that $r_\rmn{E}' \sim\,800~R_{\sun}$. These choices yield $r_\rmn{E} \sim\,2.9~\mbox{AU}$ and
$\theta_\rmn{E} \sim\,460~\mu\mbox{as}$, so that the proper motion becomes
$\mu \sim\,15~\mbox{km}\,\mbox{s}^{-1}$ and the relative lens velocity is
$v \sim\,140~\mbox{km}\,\mbox{s}^{-1}$.
With $M_V = +0.2$ and $V-I = +3.4$ for an M4 giant, an extinction of $A_I = 0.85$ yields the
baseline magnitude $I_\rmn{base} = 12.3$.

Binary lenses are likely to cause an asymmetry to the light curve, which however can be 
arbitrarily small and can even vanish for some configurations. In order to study the maximal 
impact of binarity on the measurement of limb darkening, configurations have been chosen that
preserve the symmetry. Therefore, let us assume that both lens objects have the same mass and
consider a source trajectory parallel to the line connecting their angular positions.

In general, the light received from the source is blended with additional light from other
unresolved sources (that are not affected by microlensing) or from the lens star, which is
quantified by the blend ratio $g = F_\rmn{B}/F_\rmn{S}$, where $F_\rmn{S}$ denotes the source
flux and $F_\rmn{B}$ denotes the background (blend) flux.
Since the choice of equal lens masses implies that the lens objects are M dwarfs of mass
$M/2 \sim 0.18~M_\odot$ or $M/2 \sim 0.35~M_\odot$,
their contribution to the total light falls well below the systematic error bars even at the
observed $I$-baseline. Therefore, blending is neglected with the choice $g = 0$.
Finally, a limb-darkening coefficient $\Gamma_I = 0.5$ has been adopted for both configurations.

\begin{figure*}
\includegraphics[width=84mm]{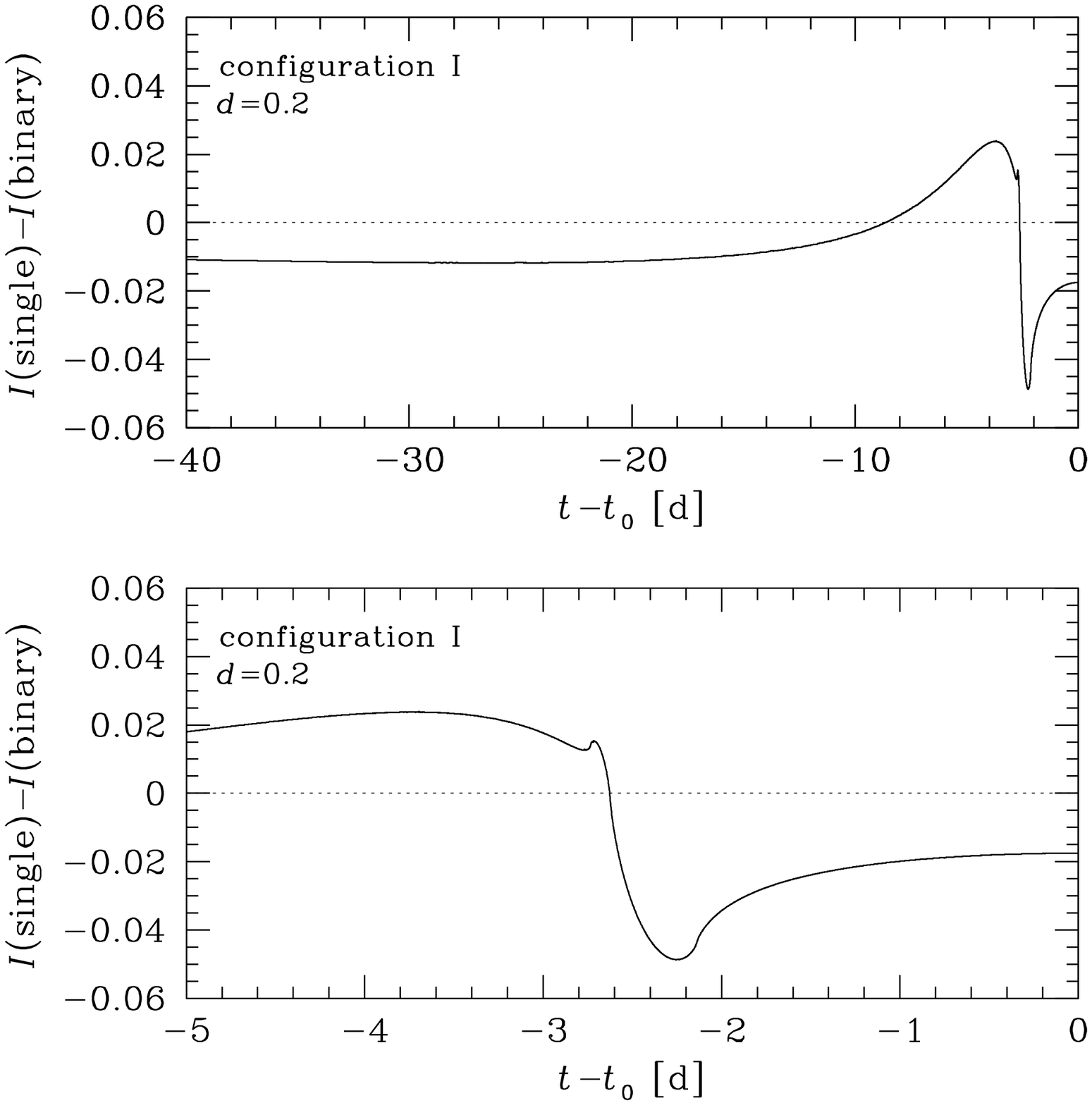}\hspace{8mm}
\includegraphics[width=84mm]{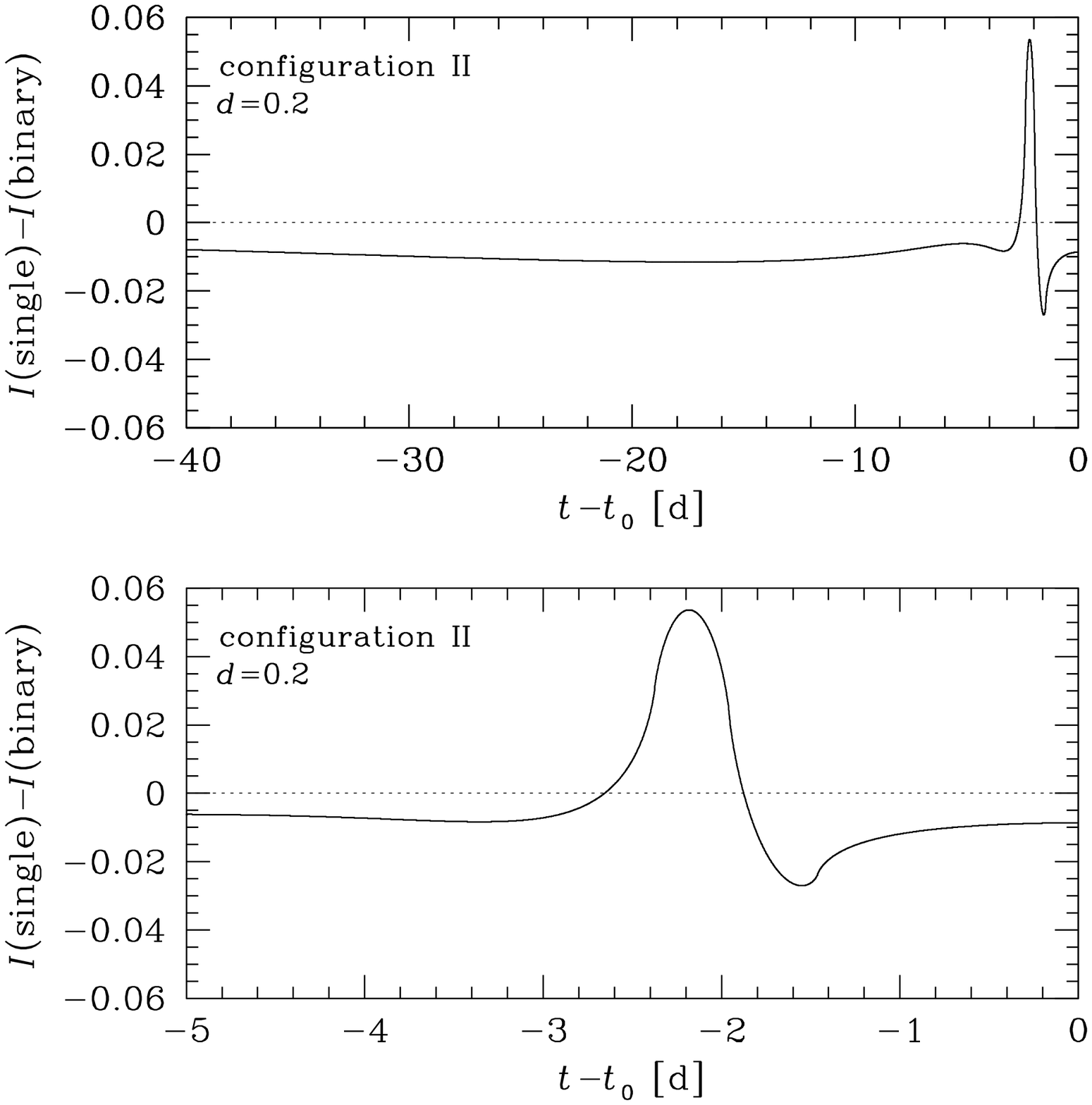}
\caption{Difference between binary-lens light curves with the angular separation parameter $d=0.2$
for configuration I (left) or
configuration II (right) 
and their single-lens counterparts.}
\label{fig:diffcurves}
\end{figure*}

The model parameters and indicative corresponding physical properties of matching lens and source
stars are summarized in Table~\ref{tab:truemodel}.
Light curves that correspond to either of the adopted configurations are shown in Fig.~\ref{fig:curves} for
a binary-lens separation parameter $d=0.2$ along with light curves that correspond to single-lens models
with otherwise identical parameters, while Fig.~\ref{fig:diffcurves} shows the difference in
magnitude between these lightcurves. Lens binarity decreases the magnification both around the
peak and in the wing region, while an increase in magnification results in regions
just before the lens center is hit by the leading limb or just after it is hit by the trailing
limb. For configuration I, these regions of increased magnification
stretch over a much larger part of the light curve than the corresponding regions for
configuration II.

\begin{figure*}
\includegraphics[width=84mm]{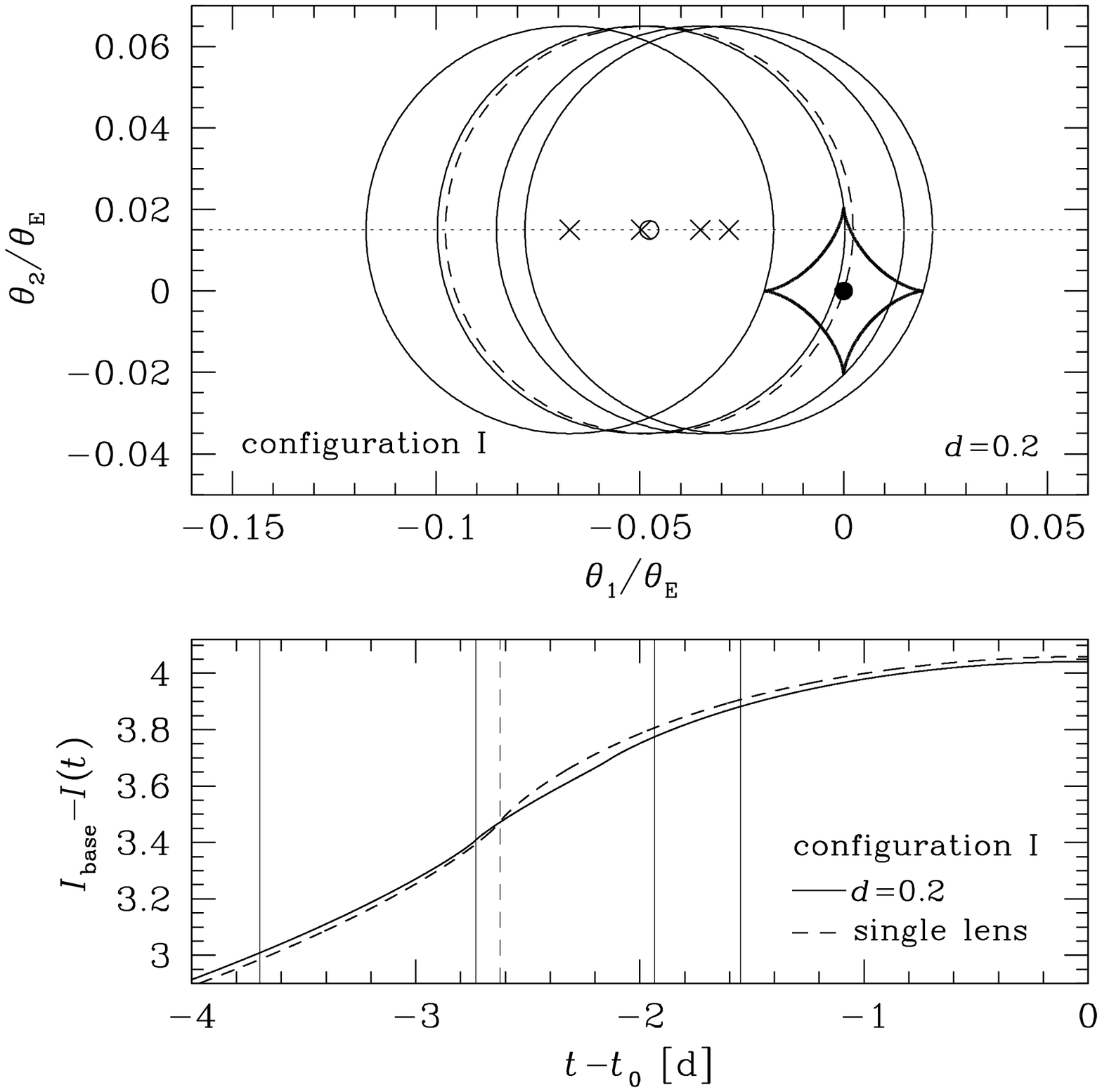}\hspace{8mm}
\includegraphics[width=84mm]{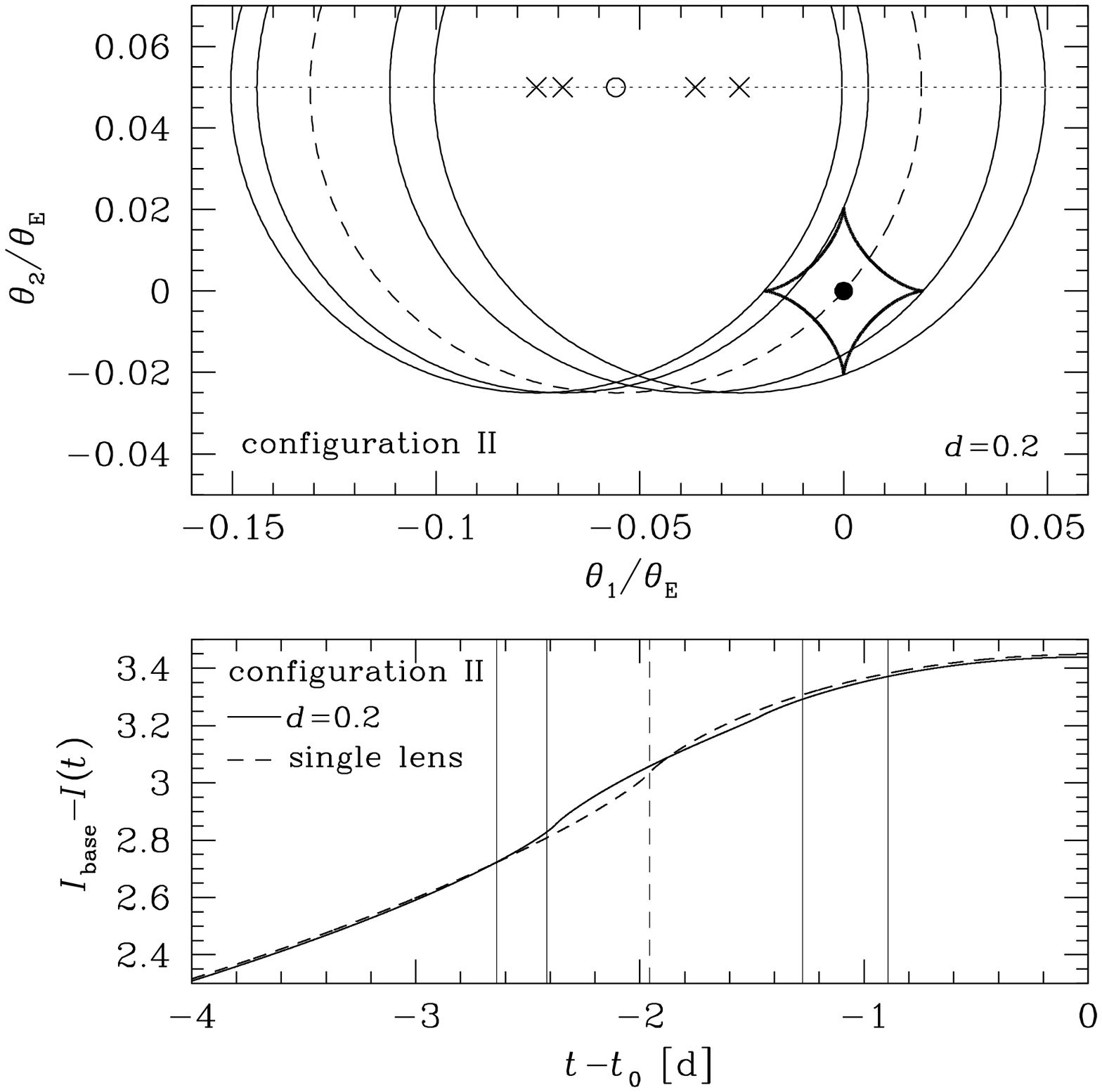}
\caption{Passages of the leading limb of the source over the four cusps for configuration I (left)
or configuration II (right) and
a binary-lens separation of $d = 0.2$. 
The upper panels show the caustic (bold line) and the lens center (filled circle) with snapshots of
the source as the trailing limb hits a cusp, as well as it touches the lens center (dashed line).
The dotted line indicates the track of the source, while diagonal crosses mark the position when a
cusp is hit, whereas the passage of the lens center is indicated by an open circle.
For configuration I, the leading limb touches the caustic for
$p_{\fss \leftarrow}^- = -0.0672$,  $p_{\fss \uparrow}^- = -0.0497$, $p_{\fss \downarrow}^- = -0.0352$, or
$p_{\fss \leftarrow}^- = -0.0282$ (in this order),
corresponding to times $t_{\fss \leftarrow}^- = -3.696~\mbox{d}$,
$t_{\fss \uparrow}^- = -2.733~\mbox{d}$, $t_{\fss \downarrow}^- = -1.935~\mbox{d}$, or
$t_{\fss \rightarrow}^- = -1.551~\mbox{d}$ relative to $t_0$, while
the lens center is hit for $p_0^- = -0.0477$, i.e.\ at $t_0^- = -2.623~\mbox{d}$.
For configuration II, the leading limb touches the cusps for
$p_{\fss \leftarrow}^- = -0.0754$,  $p_{\fss \uparrow}^- = -0.0690$,
$p_{\fss \rightarrow}^- = -0.0364$, or
$p_{\fss \downarrow}^- = -0.0255$ (in this order), and hits the lens center
for $p_0^- = -0.0559$, which corresponds to times
$t_{\fss \leftarrow}^- = -2.639~\mbox{d}$,  $t_{\fss \uparrow}^- = -2.4163~\mbox{d}$,
$t_{\fss \rightarrow}^- = -1.274~\mbox{d}$, 
$t_{\fss \downarrow}^- = -0.894~\mbox{d}$, and $t_0^- = -1.957~\mbox{d}$
relative to $t_0$.
In contrast to configuration I, the order between 
and $p_{\fss \rightarrow}^-$ and $p_{\fss \downarrow}^-$ is reversed.
 The lower
panels show the lightcurve for the binary lens (solid line) together with that for a point lens
with otherwise identical parameters (dashed line), where $t_{\fss \leftarrow}^-$,
$t_{\fss \uparrow}^-$,
$t_{\fss \downarrow}^-$, and
$t_{\fss \rightarrow}^-$
are indicated by thin solid lines, and $t_0^-$ is indicated by a thin dashed line.}
\label{fig:limbcross}
\end{figure*}

If we assume the lens center-of-mass as the origin of coordinates and the angular Einstein radius
$\theta_\rmn{E}$ as the unitlength, and if we choose the separation vector of the
binary lens objects along the $x$-axis,
the four cusps of its central
caustic are found at
\begin{equation}
\begin{array}{c}
\vec s_{\fss \leftarrow} = (s_1^-,0)\,,\quad \vec s_{\fss \rightarrow} = (s_1^+,0)\,,  \\
\vec s_{\fss \uparrow} = (\Delta s,s_2)\,,\quad \vec s_{\fss \downarrow} = (\Delta s,-s_2)\,.
\end{array}
\end{equation}
For a point source, these four cusps degenerate into the
single point $\vec s_0 = 0$, placing the caustic entry and exit at
\begin{equation}
p_0^\pm(t) = \sqrt{\rho_\star^2-u_0^2}\,.
\end{equation}

The coordinates for which the leading or trailing limb of the source 
hits a cusp then follow from the
condition $[\vec u(t)-\vec s]^2 = \rho_\star^2$, with $u(t)$ given by
Eq.~(\ref{eq:trajectory}), as
\begin{eqnarray}
p_{\fss \leftarrow}^\pm(t) & = & s_1^- \cos \alpha \pm \sqrt{\rho_\star^2-(u_0+s_1^-\sin \alpha)^2}\,,
\nonumber \\
p_{\fss \rightarrow}^\pm(t) & = & s_1^+ \cos \alpha \pm \sqrt{\rho_\star^2-(u_0+s_1^+\sin \alpha)^2}\,,
\nonumber \\
p_{\fss \uparrow}^\pm(t) & = & s_2 \sin \alpha + \Delta s \cos \alpha \pm \nonumber \\
& & \quad \pm
\sqrt{\rho_\star^2-(u_0-s_2 \cos \alpha + \Delta s \sin \alpha)^2}\,,
\nonumber \\
p_{\fss \downarrow}^\pm(t) & = & -s_2 \sin \alpha + \Delta s \cos \alpha \pm \nonumber \\
& & \quad \pm
\sqrt{\rho_\star^2-(u_0+s_2 \cos \alpha + \Delta s \sin \alpha)^2}\,.
\end{eqnarray}
For equal masses ($q=1$), $s_1^+ = -s_1^- = s_1$ and $\Delta s = 0$, while
for a point source, the four cusps degenerate into the
point $\vec s_0 = 0$, placing the caustic entry and exit at
\begin{equation}
p_0^\pm(t) = \sqrt{\rho_\star^2-u_0^2}\,.
\end{equation}

The passage of the leading stellar limb over the cusps is illustrated in Fig.~\ref{fig:limbcross}, which
shows snapshots of the source and the caustic for $d=0.2$ and
both configurations at times 
$t_{\fss \leftarrow}^-$, $t_{\fss \rightarrow}^-$, $t_{\fss \uparrow}^-$, or $t_{\fss \downarrow}^-$,
which correspond to the source positions
$p_{\fss \leftarrow}^-$, $p_{\fss \rightarrow}^-$, $p_{\fss \uparrow}^-$, or $p_{\fss \downarrow}^-$,
respectively, along with the light curve for $-4~\mbox{d} \leq t \leq 0$.
Dashed lines refer to $t_0^-$, corresponding to $p_0^-$, which is the
beginning of the caustic passage for a single point lens.

With the source size parameter $\rho_\star$ being larger for configuration II, a smaller fraction of the
source is subtended by the caustic for the same choice of the lens separation parameter $d$. A
larger impact parameter $u_0$ places the effect of binarity more towards the limb of the source.
Despite a larger source size parameter $\rho_\star$, the combination of a smaller impact parameter $u_0$ and
a smaller event time-scale $t_\rmn{E}$ leads to a  
shorter caustic passage time, which, for a single lens, is $t_0^{+}-t_0^{-} \sim 3.9~\mbox{d}$ 
for configuration II compared to 
$t_0^{+}-t_0^{-} \sim 5.2~\mbox{d}$ for configuration I. 
 
On the inclusion of a cusp, the leading limb may enter the caustic, exit the caustic, or just
touch it there so that the tangents of the circle and the merging fold lines match.
For $d=0.2$ and configuration II, the caustic is entered for $p_{\fss \leftarrow}^{-}$ and
$p_{\fss \uparrow}^{-}$, whereas it is exited for $p_{\fss \downarrow}^{-}$ and
$p_{\fss \rightarrow}^{-}$. In contrast, the caustic is exited for $p_{\fss \uparrow}^{-}$ for configuration I
and the same lens separation.
A sign change in the curvature of the light curve similar to that for a single lens near 
$t_0^{\pm}$ is seen for the binary-lens light curves near $t_{\fss \uparrow}^{\pm}$, whereas 
all other cusp intersections are not that easily identified by observable features in the light curve.

\subsection{Simulated sampled light curves}
\label{subsec:simu}

\begin{table}
\caption{Sampling intervals and number of data points for configuration I}
\label{tab:samp1}
\begin{tabular}{@{}cccc}
\hline
$t_\rmn{min}$~[d] & $t_\rmn{max}$~[d] & $\Delta t$ & $N$ \\
\hline
$0$ & $3$ & $20~\mbox{min}$ & $405$ \\
$3$ & $6$ & $30~\mbox{min}$ & $271$ \\
$6$ & $20$ & $1~\mbox{h}$ & $636$ \\
$20$ & $30$ & $2~\mbox{h}$ & $226$ \\
$30$ & $60$ & $6~\mbox{h}$ & $227$ \\
$60$ & $100$ & $12~\mbox{h}$ & $153$ \\
$100$ & $150$ & $1~\mbox{d}$ & $95$ \\
$150$ & $300$ & $2~\mbox{d}$ & $142$ \\
\hline
\end{tabular}

\medskip
With $t_0 = 0$, a constant sampling rate has been adopted on both sides of the peak
for intervals $-t_\rmn{max} \leq t \leq -t_\rmn{min}$ and $t_\rmn{min} \leq t \leq t_\rmn{max}$.
Due to the applied phase shift $f_\rmn{phase}$ and the fluctuation of the time when the 
observations are taken, $f_{\Delta t}$, the actual times of observation may fall outside the
originally designated interval. $N$ denotes the number of data points for each selected
$t_\rmn{min}$ and $t_\rmn{max}$. In total, 2155 data points have been created.
\end{table}

\begin{table}
\caption{Sampling intervals and number of data points for configuration II}
\label{tab:samp2}
\begin{tabular}{@{}cccc}
\hline
$t_\rmn{min}$~[d] & $t_\rmn{max}$~[d] & $\Delta t$ & $N$ \\
\hline
$0$ & $3$ & $20~\mbox{min}$ & $405$ \\
$3$ & $5$ & $30~\mbox{min}$ & $181$ \\
$5$ & $10$ & $1~\mbox{h}$ & $227$ \\
$10$ & $20$ & $2~\mbox{h}$ & $227$ \\
$20$ & $40$ & $4~\mbox{h}$ & $227$ \\
$40$ & $60$ & $6~\mbox{h}$ & $152$ \\
$60$ & $80$ & $12~\mbox{h}$ & $79$ \\
$80$ & $120$ & $1~\mbox{d}$ & $75$ \\
$120$ & $300$ & $2~\mbox{d}$ & $168$ \\
\hline
\end{tabular}

\medskip
Same as Table~\ref{tab:samp1} for configuration II, where a total of 1741 data points have been
created.
\end{table}

In order to see what kind of information can be extracted from sampled light curves, simulated
data sets have been created that correspond to the two chosen binary-lens configurations.
The event sampling parameters for this simulation have been chosen in analogy to
an earlier investigation of fold-caustic passages \citep{Do:FoldLD}.
For data in the range $t_\rmn{min} \leq t \leq t_{\rmn max}$, the
sampling is characterized by the sampling interval
$\Delta t$, the fluctuation $f_{\Delta t}$ 
of the time at which 
the measurement is taken, and the relative
sampling phase shift $f_\rmn{phase}$, where values of $f_{\Delta t} = 1/6$ and 
$f_\rmn{phase} = 0.2$ have been adopted here. The sampling intervals $\Delta t$ for
different regions of the light curve have been chosen
to roughly comply with the observing strategy of the PLANET microlensing follow-up campaign 
\citep{PLANET:first,PLANET:EGS} for the central region and with that of the OGLE-III survey
\citep{OGLE:general3} for the outer regions.
These are listed in Tables~\ref{tab:samp1} and~\ref{tab:samp2}.

For a reference magnitude of $m_\rmn{ref} = 16$ in $I$-band, an uncertainty of
$\sigma_\rmn{ref} = 0.015$ in the observed magnitude has been assumed, corresponding to
a $\sim\,1.5\,$\% relative uncertainty in the measured flux. For other magnitudes, it is 
assumed that the photometric measurement follows Poisson statistics, so that
its uncertainty is proportional
to the square-root of the observed flux. This uncertainty
has been smeared with a relative standard deviation of $f_\sigma = 0.125$.
The resulting photometric error bar $\sigma_\rmn{phot}$ yet
does not represent the total measurement
uncertainty. Instead, a systematic error $\sigma_0$ is added in quadrature,
which becomes dominant as the target gets bright. For the systematic error, three different
values $\sigma_0 = 0.3\,\%$, $\sigma_0 = 0.5\,\%$, or $\sigma_0 = 1\,\%$ have been used.

\begin{figure}
\includegraphics[width=84mm]{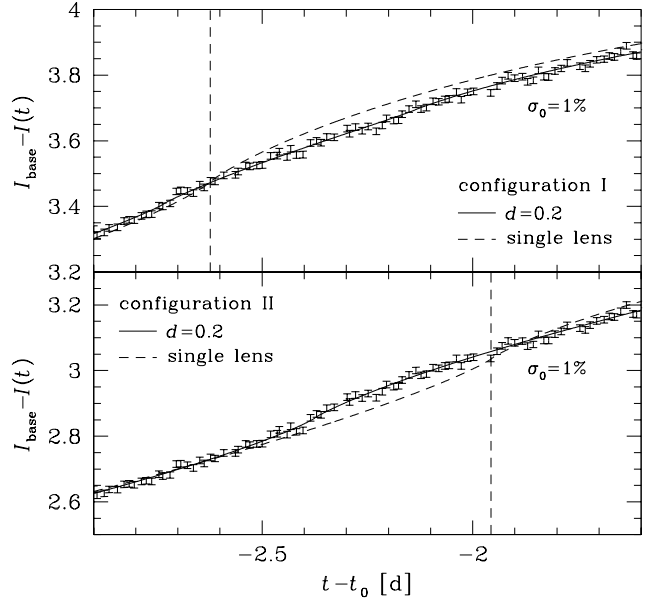}
\caption{Binary-lens light curves (solid line) and simulated data for $\sigma_0 = 1~\%$ for
both configurations and
a lens separation parameter $d=0.2$. Also shown are light curves for corresponding single-lens
models with otherwise identical parameters (dashed line). The time $t_0^-$ at which 
the leading limb hits the single-lens point caustic is indicated by a vertical thin dashed line.}
\label{fig:limbdat}
\end{figure}

All the above choices mean that the creation of synthetic
data sets ($t^{(i)}$, $m^{(i)}$, $\sigma^{(i)}$) 
follows the rules
\begin{eqnarray} 
t^{(i)} & = & \mathcal{N}\left[t_\rmn{min}+(i+f_\rmn{phase})\,\Delta t,f_{\Delta t}\,\Delta t\right]\,,\\
{\overline \sigma}^{(i)} & = & \sigma_\rmn{ref}\,10^{0.2[m(t^{(i)})-m_\rmn{ref}]}\,, \\
\sigma_\rmn{phot}^{(i)} & = & \mathcal{N}\left({\overline \sigma}^{(i)},f_\rmn\sigma\,
{\overline \sigma}^{(i)}\right)\,, \\
\sigma^{(i)} & = & \sqrt{\left[\sigma_\rmn{phot}^{(i)}\right]^2 + \sigma_0^2} \,, \\
m^{(i)} & = & \mathcal{N}\left(m(t^{(i)}),\sigma^{(i)}\right)\,,
\end{eqnarray}
where $\mathcal{N}(\mu,\sigma)$ denotes a value drawn randomly from a normal distribution with mean
$\mu$ and standard deviation $\sigma$, $m(t^{(i)})$ denotes the 
magnitude at time $t^{(i)}$ for the adopted model parameters,
and $i \in [0,(t_\rmn{max}-t_\rmn{min})/(\Delta t)]$.

In order to simulate possible losses in the data acquisition 
due to observing conditions
or technical problems, data points have been removed from the
dataset at random with a probability of $p_\rmn{loss} = 5\,$\%.

In fact, a round-the-clock coverage (without daylight gaps)
requires more than one telescope. Nevertheless, the effects of additional free parameters 
(mainly baselines magnitudes) and different telescope characteristics
for multi-site observations are neglected in favour of simplicity, and the data is treated such
as resulting from a single telescope.

The quality of coverage of the data set is demonstrated in Fig.~\ref{fig:limbdat}, which
shows the simulated data points and the theoretical light curves
for $d=0.2$ and $\sigma_0 = 1~\,\%$ for both binary-lens
configurations around the contact of the leading limb with the lens center. To indicate
the influence of lens binarity, the corresponding single-lens light curves are also
shown.

\subsection{Best-matching single-lens models}
\label{subsec:results}

\begin{table*}
\begin{minipage}{176mm}
\caption{Best-matching single-lens models to configuration I binary-lens models with
different angular lens separations for different systematic errors.}
\label{tab:conf1}
\begin{tabular}{@{}l|ccc|ccc|ccc}
\hline
 & \multicolumn{3}{|c|}{$\sigma_0 = 0.3\,$\%} & \multicolumn{3}{|c|}{$\sigma_0 = 0.5\,$\%} &
 \multicolumn{3}{|c|}{$\sigma_0 = 1\,$\%} \\
 & $d=0.2$ & $d=0.15$ & $d=0.1$ & $d = 0.2$ & $d=0.15$ & $d=0.1$ & $d = 0.2$ & $d=0.15$ & $d=0.1$\\
\hline
$\chi^2_\rmn{min}$ & $18054.9$ & $4337.6$ & $2235.6$ & $8873.9$ & $2969.9$ & $2097.89$ 
& $3895.1$ & $2267.8$ & $2037.6$ \\
d.o.f. & $2149$ & $2149$ & $2149$ &  $2149$ & $2149$ & $2149$ & $2149$ & $2149$ & $2149$ \\
$\chi^2_\rmn{min}$/d.o.f. & $8.40$ & $2.02$ & $1.04$ & $4.13$ & $1.38$ & $0.976$
& $1.81$ & $1.06$ & $0.948$ \\
$\sqrt{2 \chi^2_\rmn{min}}-\sqrt{2n-1}$ & $124.5$ & $27.6$ & $1.32$ & $67.7$ & $11.5$ & $-0.78$ 
& $22.7$ & $1.80$ & $-1.72$ \\
$P(\chi^2 \geq \chi^2_\rmn{min})$ & $\la 10^{-3366}$ & $\la 10^{-167}$ & $0.094$ &
$\la 10^{-996}$ & $\la 10^{-30}$ & $0.781$ & $\la 10^{-113}$ & $0.037$ & $0.957$ \\
\hline
$t_\rmn{E}$ & $55.44$ & $54.99$ & $54.93$ & $55.28$ & $54.94$ & $54.91$ 
& $55.18$ & $54.91$ & $54.91$ \\
$u_0$ & $0.0157$ &  $0.0149$ & $0.0148$ & $0.0155$ & $0.0148$ & $0.0147$ 
& $0.0153$ & $0.0148$ & $0.0147$ \\
$\rho_\ast$ & $0.0541$ & $ 0.0518$ & $0.0503$ & $0.0542$ & $0.0519$ & $0.0503$ &
 $0.0543$ & $0.0520$ & $0.0503$ \\
$I_\rmn{base}$ & $13.609$ & $13.604$ & $13.601$ & $13.608$ & $13.604$ & $13.602$ &
$13.607$ & $13.603$ & $13.602$ \\
$\Gamma_I$ & $1.000$ & $0.746$ & $0.538$ & $1.000$ & $0.751$ & $0.538$ &
$1.000$ & $0.751$ & $0.537$
\\
\hline
\end{tabular}

\medskip
The complete data set has been included. The 'true' binary-lens parameters are 
shown in Table~\ref{tab:truemodel}. The blending parameter has been fixed to $g=0$, whereas
$t_0$ has been allowed to vary, yielding $|t_0| \la 90~\mbox{s}$ in all cases.
The quantity $\sqrt{2 \chi^2_\rmn{min}}-\sqrt{2n-1}$ is
a measure of the goodness-of-fit as a characteristic of a $\chi^2$-test yielding the
equivalent deviation from the mean in units of the standard deviation for a Gaussian distribution,
where $n$ is the number of degrees of freedom (d.o.f.), which is the number of data points reduced by the number
of free model parameters, while $P(\chi^2 \geq \chi^2_\rmn{min})$ denotes the associated 
probability.
\end{minipage}
\end{table*}

\begin{table*}
\begin{minipage}{176mm}
\caption{Best-matching single-lens models to configuration II binary-lens models with
different angular lens separations for different systematic errors.}
\label{tab:conf2}
\begin{tabular}{@{}l|ccc|ccc|ccc}
\hline
 & \multicolumn{3}{|c|}{$\sigma_0 = 0.3\,$\%} & \multicolumn{3}{|c|}{$\sigma_0 = 0.5\,$\%} &
 \multicolumn{3}{|c|}{$\sigma_0 = 1\,$\%} \\
 & $d=0.2$ & $d=0.15$ & $d=0.1$ & $d = 0.2$ & $d=0.15$ & $d=0.1$ & $d = 0.2$ & $d=0.15$ & $d=0.1$\\
\hline
$\chi^2_\rmn{min}$ & $5670.3$ & $2371.2$ & $1791.5$ & $3161.4$ & $1895.0$ & $1666.8$ &
$2001.6$ & $1668.7$ & $1605.0$ \\
d.o.f. & $1735$ & $1735$ & $1735$ & $1735$ & $1735$ & $1735$ & $1735$ & $1735$ & $1735$  \\
$\chi^2_\rmn{min}$/d.o.f. & $3.27$ & $1.37$ & $1.03$ & $1.82$ & $1.09$ & $0.961$ &
$1.15$ & $0.962$ & $0.925$ \\ 
$\sqrt{2 \chi^2_\rmn{min}}-\sqrt{2n-1}$ & $47.6$ & $10.0$ & $0.96$ & $20.6$ & $2.66$ & $-1.16$ &
$4.4$ & $-1.13$ & $-2.24$ \\ 
$P(\chi^2 \geq \chi^2_\rmn{min})$ &$\la 10^{-494}$ & $\la 10^{-23}$ & $0.168$ &
$\la 10^{-94}$ & $0.004$ & $0.877$ & $\la 10^{-5}$ & $0.871$ & $0.988$ \\ 
\hline
$t_\rmn{E}$ & $34.62$ & $34.76$ & $34.91$ & $34.62$ & $34.77$ & $34.91$ &
$34.64$ & $34.78$ & $34.93$ \\ 
$u_0$ & $0.0489$ & $0.0491$ & $0.0497$ & $0.0489$ & $0.0491$ & $0.0497$ 
& $0,0489$ & $0,0491$ & $0,0497$ \\ 
$\rho_\ast$ & $0.0817$ & $0.0795$ & $0.0758$ & $0.0817$ & $0.0796$ & $0.0758$ &
$0.0816$ & $0.0796$ & $0.0757$ \\ 
$I_\rmn{base}$ & $12.302$ & $12.301$ & $12.301$ & $12.302$ & $12.301$ & $12.301$ &
$12.302$ & $12.302$ & $12.301$ \\ 
$\Gamma_I$ & $1.000$ & $0.984$ & $0.607$ & $1.000$ & $0.997$ & $0.605$ &
$1.000$ & $1.000$ & $0.599$ 
\\
\hline
\end{tabular}

\medskip
The same quantities as in Table~\ref{tab:conf1} are displayed. Again, all data points have been
included in the fits and the 'true' binary-lens parameters are listed in Table~\ref{tab:truemodel}.
The blending parameter has been fixed to $g=0$, while $t_0$ was allowed to vary, where
$|t_0| \la 90~\mbox{s}$ resulted in all cases. 
\end{minipage}
\end{table*}

Let us now assume a single lens and determine the corresponding model parameters by
means of fits to the simulated data sets that correspond to the two binary-lens configurations.
We will then see whether statistical tests suggest to accept the single-lens model and 
how well the true limb-darkening coefficient $\Gamma$ and the time-scale $t_\star$, which
yields the proper motion as $\mu = \theta_\star/t_\star$, are reproduced. Any significant
offsets for acceptable single-lens models will limit the accuracy to which these paramters are
determined on the assumption of such a model.

Tables~\ref{tab:conf1} and~\ref{tab:conf2} list the obtained model parameters and the results
of $\chi^2$ tests for both binary-lens configurations and the different lens separations $d$ and
systematic errors $\sigma_0$. 
For configuration I, $\chi^2$ tests recommend to accept the single-lens
model for $d=0.1$ and all applied systematic errors, and marginally for $d = 0.15$ and
$\sigma = 1\,\%$. The limb-darkening coefficient $\Gamma_I$ 
turns out to be $\sim 8\,\%$
larger for $d=0.1$, $\sim 50\,\%$ larger for $d=0.15$, and tends to its maximal value
$\Gamma_I = 1$ for $d=0.2$. Compared to $\Gamma_I$, the relative offsets on $t_\star$ are
much smaller, namely $\sim 0.5\,\%$ for $d=0.1$, $\sim 4\,\%$ for $d=0.15$, and
$\sim 10\,\%$ for $d=0.2$, usually below the uncertainty in the measurement of $\theta_\star$.

Similar to configuration I, the acceptance of the single-lens model for $d=0.1$ and all
applied systematic errors
as well as for $d=0.15$ and $\sigma = 1\,\%$ is recommended by $\chi^2$ tests for configuration II.
A larger goodness-of-fit than for configuration I comes along with a larger offset in
$\Gamma_I$, namely $\sim 20\,\%$ for $d=0.1$, while
$\Gamma_I \sim 1$ is already reached for $d=0.15$. The time-scale $t_\star$ is found to
be shifted by $\sim 0.8\,\%$ for $d=0.1$, $\sim 5\,\%$ for $d=0.15$, and
$\sim 8\,\%$ for $d=0.2$,

\begin{figure*}
\includegraphics[width=84mm]{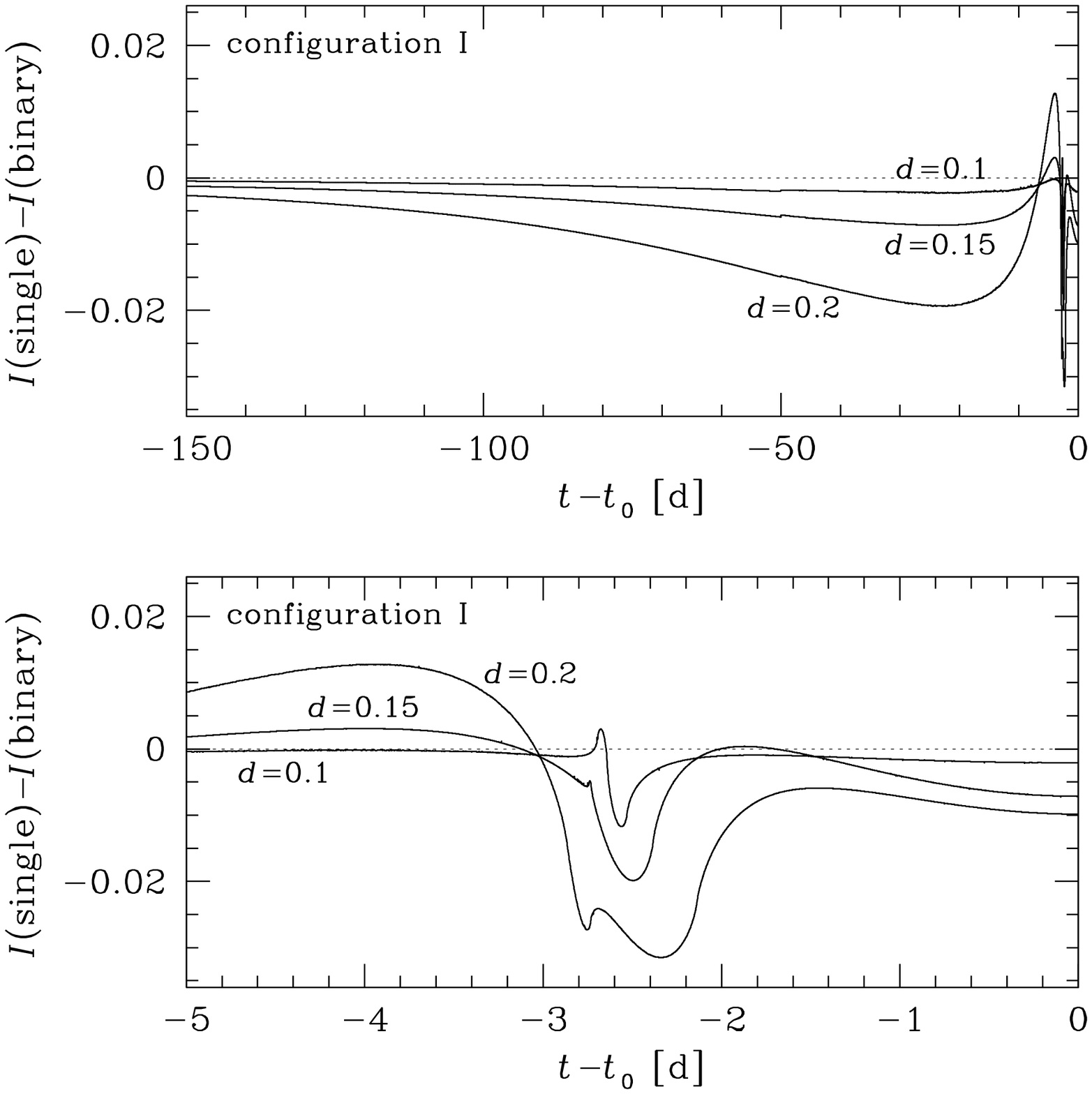}\hspace{8mm}
\includegraphics[width=84mm]{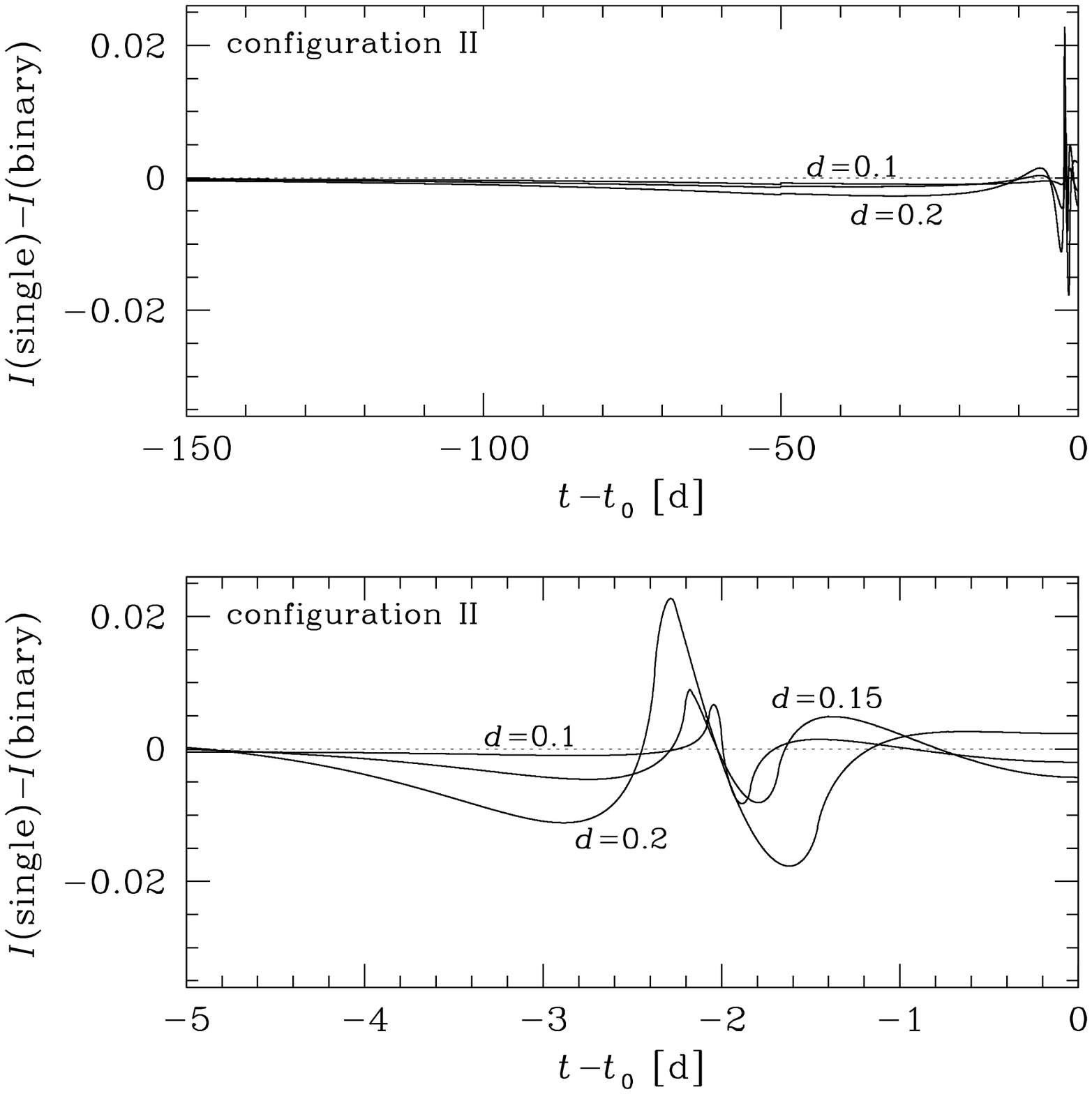}
\caption{Differences in the light curve between a binary-lens model and a best-matching single
lens model. The binary-lens models shown correspond to configuration I (left) or
configuration II (right) and separations of $d=0.2$, $d=0.15$, and $d=0.1$,
while the single-lens model is the best fit for $\sigma_0 = 0.3~\%$ as listed in
Table~\ref{tab:conf1} or Table~\ref{tab:conf2}, respectively. The upper panels show the full wing
 of the light curve, whereas the lower panels show only the peak region.}
\label{fig:confdel}
\end{figure*}

The difference between the light curves for best-matching single-lens models and the 
'true' binary-lens configuration for the three lens separations
$d=0.2$, $d=0.15$, or $d=0.1$ is shown in Fig.~\ref{fig:confdel}. As Tables~\ref{tab:conf1}
and~\ref{tab:conf2} show, the model parameters differ only very slightly 
for the different choices of the
systematic error $\sigma_0$, so that the corresponding light curves are practically identical.
For Fig.~\ref{fig:confdel}, $\sigma_0 = 0.3~\%$ has been chosen.
For configuration I, some of the discrepancy is moved into the wing region of the light curve, 
where differences of up to $2\,\%$ (for $d=0.2$), $0.7\,\%$ (for $d=0.15$) or 
$0.2\,\%$ (for $d=0.1$) occur. In contrast, for configuration II, the maximal differences in
the wing region are only $0.25\,\%$ (for $d=0.2$), $0.13\,\%$ (for $d=0.15$) or 
$0.1\,\%$ (for $d=0.1$). This reflects the fact that the region for which the binary-lens
magnification exceeds that of a single lens extends significantly outside the region where
the stellar limb touches the caustic for configuration I, while this is not the case for
configuration II. The most significant deviations occur around times when the stellar limb touches
the caustic, which are however below $3\,\%$ (for $d=0.2$), $2\,\%$ (for $d=0.15$) or 
$1.2\,\%$ (for $d=0.1$) for configuration I, and
below $2.5\,\%$ (for $d=0.2$), $1\,\%$ (for $d=0.15$) or 
$0.8\,\%$ (for $d=0.1$) for configuration II.

Since it is quite challenging to achieve photometric uncertainties below the $1\,\%$-level,
one should consider the effect of lens binarity as a serious problem for measuring limb-darkening
coefficients from close-impact microlensing events and be quite careful with results obtained under
the assumption of a single lens.

An additional complication arises from the fact that photometric error bars reported by
reduction algorithms such as DoPhot \citep{DoPhot} tend to underestimate the true
photometric uncertainty for microlensing observations, which are plagued by varying
observing conditions during the course of an event. Apart from the systematic error that is
taken into account for the simulated data set, underestimates of 10--20~\% occur frequently
\citep[e.g.][]{OGLEerr,PLANET:first,Tsapras}, so that
the ratio $\chi^2/\mathrm{d.o.f.}$ is not unlikely to exceed 1.4.
This strongly limits the significance of the $\chi^2$ test as a measure of the goodness-of-fit.

If one accepts an underestimate by $20~\%$, the single-lens model for the configuration I
binary-lens data with $d=0.15$ and $\sigma = 0.5\,\%$ becomes acceptable, while for configuration
II, single-lens models look acceptable for $d=0.15$ and all applied systematic errors as well
as for $d=0.2$ and $\sigma = 1\,\%$.

\begin{table*}
\begin{minipage}{176mm}
\caption{Best-matching single-lens models to the peak region of binary-lens light curves for
both discussed configurations and different angular lens separations.}
\label{tab:peak}
\begin{tabular}{@{}l|ccc|ccc}
\hline
 & \multicolumn{6}{c}{$\sigma_0 = 0.3\,\%$} \\
 & \multicolumn{3}{c}{configuration I} & \multicolumn{3}{c}{configuration II} \\
 & $d=0.2$ & $d=0.15$ & $d=0.1$ & $d = 0.2$ & $d=0.15$ & $d=0.1$ \\
\hline
$\chi^2_\rmn{min}$ & $1121.4$ & $926.8$ & $461.9$ & $3468.21$ & $948.98$ & $572.34$ \\
d.o.f.  & $398$ & $398$ & $398$ & $398$ & $398$ & $398$\\
$\chi^2_\rmn{min}$/d.o.f.  & $2.82$ & $2.32$ & $1.16$ & $8.7$ & $2.38$ & $1.44$ \\
$\sqrt{2 \chi^2_\rmn{min}}-\sqrt{2n-1}$ &  $19.2$ & $14.9$ & $2.20$ & $55.1$ & $15.4$ & $5.6$ \\
$P(\chi^2 \geq \chi^2_\rmn{min})$ & $\la 10^{-81}$ & $\la 10^{-49}$ & $0.015$ &
$\la 10^{-661}$ & $\la 10^{-53}$ & $\la 10^{-8}$ \\
\hline
$t_\rmn{E}$ & $90.68$ & $58.51$ & $56.01$ & $34.02$ & $34.47$ & $34.92$ \\
$u_0$ & $0.0140$ & $0.0181$ & $0.0159$ & $0.0478$ & $0.0487$ & $0.497$ \\
$\rho_\ast$ &  $0.0332$ & $0.0487$ & $0.0495$ & $0.0829$ & $0.0802$ & $0.0758$ \\
$g$  & $0.571$ &$0.031$ & $0.011$ & $0.0034$ & $0.0014$ & $0.0005$ \\
$\Gamma_I$ & $1.000$ & $0.663$ & $0.531$ & $1.000$ & $1.000$ & $0.605$
\\
\hline
\end{tabular}

\medskip
Results of fits of single-lens models to the peak region
($-3~\mbox{d} \leq t \leq 3~\mbox{d}$) of the simulated binary-lens data sets for both
configurations for
a systematic error $\sigma_0 = 0.3~\%$ and different angular lens separations. The parameters
of the underlying binary-lens models can be found in
Table~\ref{tab:truemodel}. In contrast to the fits that include
the full data set, the baseline magnitude $I_\rmn{base}$ has been fixed, whereas the blending parameter
$g$ has been allowed to vary. Again, the free parameter $t_0$ fulfilled $|t_0| \la 90~\mbox{s}$ in all cases. 
\end{minipage}
\end{table*}

In addition to fits making use of the complete set of data points, best-matching single-lens models
for the peak region of the lightcurve, defined as $-3~\mbox{d} \leq t \leq 3~\mbox{d}$, have
been obtained for $\sigma_0 = 0.3~\,\%$, for which the model parameters and the
result of $\chi^2$ tests are
displayed in Table~\ref{tab:peak}. In contrast to the fits recognizing the full data set,
the baseline
$I_\rmn{base}$ has been fixed to its 'true' value, while the blending parameter $g$ has been
allowed to vary.
As one could have expected from the larger discrepancies being
attributed to the wing regions of the light curves for configuration I, the model parameters for
the fits restricted to the peak region deviate more strongly from those obtained when the
full data set is
considered. In order to adjust to the optimal model,
the blending parameter $g$ has assumed 
a significant non-zero value. 
For all selected binary-lens separations $d$ for configuration II and for $d=0.1$ for 
configuration I, the goodness-of-fit resulting from the $\chi^2$ test for the peak region model and
data is worse than for the models and data for the full light curve, so that a rejection of
all considered single-lens models is indicated.  However, if one accounts for a possible $20~\%$ 
increase in the size of the photometric errors, the models with $d=0.1$ for both configurations
still survive. In any case, the $d=0.1$ models are not recommended for rejection
for the larger systematic errors $\sigma_0 = 0.5~\%$ or $\sigma_0 = 1\,\%$ 
by means of a $\chi^2$ test over the peak region.

\begin{figure}
\includegraphics[width=84mm]{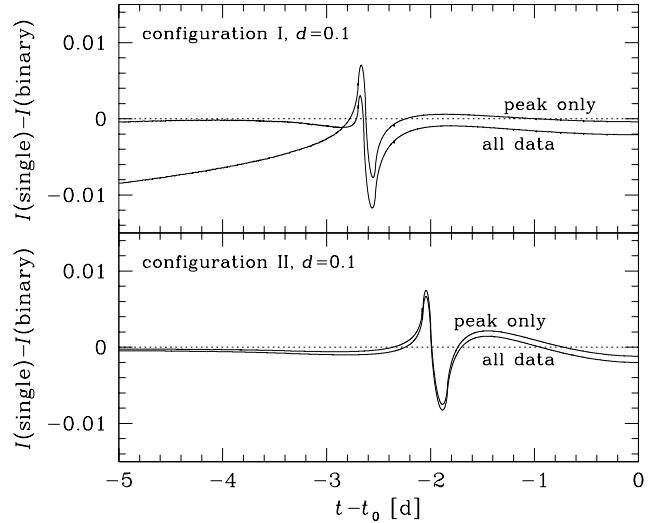}
\caption{Differences in the light curve between a binary-lens model and a best-matching single
lens model. Comparison between fits for the full data set (bold lines) and those where only the peak 
region ($-3~\mbox{d} \leq t \leq 3~\mbox{d}$) has been included (thin lines)
with $\sigma_0 = 0.3~\%$ for both discussed configurations and a binary separation $d = 0.1$.}
\label{fig:peakdel}
\end{figure}

Figure~\ref{fig:peakdel} shows a comparison of the differences between the binary-lens and the
single-lens light curves of fits involving the full data set and those restricted
to the peak region for
a binary-lens separation parameter $d=0.1$. While the differences between these two types of
fits are negligible for configuration II, the apparent
differences for configuration I demonstrate the
amount of information contained in the region before the first and after the last contact of the
stellar limb with the caustic. For configuration I, the maximal deviation is reduced from
$1.2\,\%$ to $0.8\,\%$ for the peak-only fits in the peak region at the cost of a deviation
of up to  $1.4\,\%$ outside compared to $0.2\,\%$ when the full data set is considered.

\begin{table*}
\begin{minipage}{176mm}
\caption{Results of run tests for the best-matching single-lens models for configuration I.}
\label{tab:run1}
\begin{tabular}{@{}cl|ccc|ccc|ccc}
\hline
 & & \multicolumn{3}{|c|}{$\sigma_0 = 0.3\,$\%} & \multicolumn{3}{|c|}{$\sigma_0 = 0.5\,$\%} &
 \multicolumn{3}{|c|}{$\sigma_0 = 1\,$\%} \\
 & & $d=0.2$ & $d=0.15$ & $d=0.1$ & $d = 0.2$ & $d=0.15$ & $d=0.1$ & $d = 0.2$ & $d=0.15$ & $d=0.1$\\
\hline
& $N_+$ & $911$ & $1013$ & $1051$ & $935$ & $1034$ & $1063$ & $1016$ & $1053$ & $1071$ \\
& $N_-$ & $1244$ & $1142$ & $1104$ & $1220$ & $1121$ & $1092$ & $1139$ & $1102$ & $1084$ \\
& $\mathcal{E}(n_\rmn{r})$ & $1052.8$ & $1074.6$ & $1077.8$ &
$1059.7$ & $1076.7$ & $1078.3$ & $1075.0$ & $1077.9$ & $1078.5$ \\
\raisebox{1.5ex}[-1.5ex]{$N = 2155$} & $\sigma(n_\rmn{r})$ & $22.7$ & $23.1$ & $23.2$ &
$22.8$ & $23.2$ & $23.2$ & $23.1$ & $23.2$ & $23.2$ \\
\raisebox{1.5ex}[-1.5ex]{(all data)} & $n_\rmn{r}^\rmn{obs}$ & 
$365$ & $774$ & $1064$ & $523$ & $928$ & $1086$ & $816$ & $1032$ & $1122$ \\
& $\delta$ & $30.4$ & $13.0$ & $0.60$ & $23.5$ & $6.4$ & $-0.33$ & $11.2$ & $1.98$ & $-1.88$ \\
 & $P_\rmn{r}$ & $\la 10^{-202}$ & $\la 10^{-38}$ & $0.275$ & $\la 10^{-122}$ & $\la 10^{-10}$ &
 $0.630$ & $\la 10^{-28}$ & $0.024$ & $0.970$ \\
\hline
& $N_+$ & $187$ & $183$ & $197$ & $169$ & $179$ & $195$ & $164$ & $186$ & $199$ \\
& $N_-$ & $217$ & $221$ & $207$ & $235$ & $225$ & $209$ & $240$ & $218$ & $205$ \\
& $\mathcal{E}(n_\rmn{r})$ & $201.9$ & $201.2$ & $202.9$ &
$197.6$ & $200.4$ & $202.8$ & $195.9$ & $201.7$ & $203.0$ \\
\raisebox{1.5ex}[-1.5ex]{$N = 404$} & $\sigma(n_\rmn{r})$ & $10.0$ & $10.0$ & $10.0$ &
$9.8$ & $9.9$ & $10.0$ & $9.7$ & $10.0$ & $10.4$ \\
\raisebox{1.5ex}[-1.5ex]{(peak only)} & $n_\rmn{r}^\rmn{obs}$ &
$111$ & $120$ & $190$ & $123$ & $152$ & $198$ & $152$ & $178$ & $204$ \\
& $\delta$ & $9.1$ & $8.2$ & $1.28$ & $7.6$ & $4.9$ & $0.48$ & $4.5$ & $2.4$ & $-0.10$ \\
 & $P_\rmn{r}$ & $\la 10^{-19}$ & $\la 10^{-15}$ & $0.100$ & $\la 10^{-14}$ &
 $\la 10^{-6}$ & $0.317$ & $\la 10^{-5}$ & $0.009$ & $0.541$ \\
 \hline
\end{tabular}

\medskip
For the fits of single-lens models to the full data set corresponding to configuration I
and different values of the lens separation $d$ and the systematic error $\sigma_0$
whose parameters are displayed
in Table~\ref{tab:conf1},
run tests over the whole data set or the 
peak region ($-3~\mbox{d} \leq t \leq 3~\mbox{d}$) revealed 
$N_{+}$ positive and $N_{-}$ negative residuals, where 'positive' means that the observed 
magnification exceeds the theoretical one. $\mathcal{E}(n_\rmn{r})$ denotes the expected 
number of runs and $\sigma(n_\rmn{r})$ denotes the corresponding standard deviation. 
For $n_\rmn{r}^{\rmn{obs}}$ runs being found, the deviation in units of $\sigma(n_\rmn{r})$
becomes $\delta = [\mathcal{E}(n_\rmn{r})-n_\rmn{r}^{\rmn{obs}}]/
\sigma(n_\rmn{r})$, which corresponds to a probability $P_\rmn{r} = P(n_\rmn{r} \leq 
n_\rmn{r}^\rmn{obs})$.  
\end{minipage}
\end{table*}

\begin{table*}
\begin{minipage}{176mm}
\caption{Results of run tests for the best-matching single-lens models for configuration II.}
\label{tab:run2}
\begin{tabular}{@{}cl|ccc|ccc|ccc}
\hline
 & & \multicolumn{3}{|c|}{$\sigma_0 = 0.3\,$\%} & \multicolumn{3}{|c|}{$\sigma_0 = 0.5\,$\%} &
 \multicolumn{3}{|c|}{$\sigma_0 = 1\,$\%} \\
 & & $d=0.2$ & $d=0.15$ & $d=0.1$ & $d = 0.2$ & $d=0.15$ & $d=0.1$ & $d = 0.2$ & $d=0.15$ & $d=0.1$\\
\hline
& $N_+$ & $888$ & $852$ & $846$ & $869$ & $850$ & $841$ & $856$ & $856$ & $839$ \\
& $N_-$ & $853$ & $889$ & $905$ & $872$ & $891$ & $900$ & $885$ & $885$ & $902$ \\
& $\mathcal{E}(n_\rmn{r})$ & $871.2$ & $871.1$ & $870.8$ & $871.5$ & $871.0$ & $870.5$ &
$871.3$ & $871.3$ & $870.4$ \\
\raisebox{1.5ex}[-1.5ex]{$N = 1741$} & $\sigma(n_\rmn{r})$ & $20.9$ & $20.9$ & $20.8$ &
$20.9$ & $20.9$ & $20.8$ & $20.9$ & $20.9$ & $20.8$ \\
\raisebox{1.5ex}[-1.5ex]{(all data)} & $n_\rmn{r}^\rmn{obs}$ &
$665$ & $776$ & $852$ & $729$ & $828$ & $884$ & $810$ & $862$ & $884$ \\
& $\delta$ & $9.9$ & $4.6$ & $0.90$ & $6.8$ & $2.1$ & $-0.65$ & $2.9$ & $0.44$ & $-0.66$ \\
 & $P_\rmn{r}$ & $\la 10^{-22}$ & $\la 10^{-5}$ & $0.183$ & $\la 10^{-11}$ &
 $0.020$ & $0.742$ & $0.002$ & $0.329$ & $0.744$ \\
\hline
& $N_+$ & $218$ & $189$ & $200$ & $203$ & $187$ & $200$ & $196$ & $198$ & $196$ \\
& $N_-$ & $186$ & $215$ & $204$ & $201$ & $217$ & $204$ & $208$ & $206$ & $208$ \\
& $\mathcal{E}(n_\rmn{r})$ & $201.7$ & $202.2$ & $203.0$ & $203.0$ & $201.9$ & $203.0$ &
$202.8$ & $202.9$ & $202.8$ \\
\raisebox{1.5ex}[-1.5ex]{$N = 404$} & $\sigma(n_\rmn{r})$ & $10.0$ & $10.0$ & $10.0$ &
$10.0$ & $10.0$ & $10.0$ & $10.0$ & $10.0$ & $10.0$ \\
\raisebox{1.5ex}[-1.5ex]{(peak only)} & $n_\rmn{r}^\rmn{obs}$ &
$51$ & $107$ & $166$ & $93$ & $147$ & $190$ & $141$ & $174$ & $194$ \\
& $\delta$ & $15.1$ & $9.5$ & $3.7$ & $11.0$ & $5.5$ & $1.29$ & $6.2$ & $2.9$ & $0.88$ \\
 & $P_\rmn{r}$ & $\la 10^{-51}$ & $\la 10^{-21}$ & $\la 10^{-4}$ & $\la 10^{-27}$ &
 $\la 10^{-7}$ & $0.098$ & $\la 10^{-9}$ & $0.002$ & $0.189$ \\
 \hline
\end{tabular}

\medskip
Results of run tests corresponding to fits of single-lens models to the full data set corresponding to configuration II 
and different values of the lens separation $d$ and the systematic error $\sigma_0$
whose parameters are displayed
in Table~\ref{tab:conf2}, where either
the whole data set or the peak region ($-3~\mbox{d} \leq t \leq 3~\mbox{d}$) has been used.
The quantities shown are the same as in Table~\ref{tab:run1}. 
\end{minipage}
\end{table*}

Given the problems in rejecting single-lens models by means of a $\chi^2$ test,
it seems to be a good idea to look for characteristic deviation patterns in order to
decide on whether a model is in agreement with the observed data.
An underestimate of error bars is not a problem for the
assessment of a run test, which recognizes the sign of the residuals only, but not
their size. In this sense, it is complementary to the $\chi^2$ test, which is sensitive to 
the absolute value of the residuals, but blind to their signs.
Although the obtained $\chi^2_\rmn{min}$
may look appropriate, a model is not acceptable if it fails a run test.
Let a 'run' be defined as the longest contiguous sequence of
residuals with the same sign, and let $N$ denote the total number of 
data points, $N_{+}$ the number of points with positive residuals, and
$N_{-}$ the number of points with negative residuals.
For $N > 10$, the distribution of the number of runs $n_\rmn{r}$ can 
be fairly approximated by a normal distribution with the expectation value
\begin{equation}
\mathcal{E}(n_\rmn{r})  =  1+\frac{2 N_{+} N_{-}}{N}
\end{equation}
and the standard deviation
\begin{equation}
\sigma(n_\rmn{r})  =  \sqrt{2 N_{+} N_{-} 
\frac{2 N_{+} N_{-} - N}{N^2 (N-1)}}\,.
\end{equation}
The goodness-of-fit can then be measured by the probability $P_\rmn{r} = P(n_\rmn{r}
\leq n_\rmn{r}^\rmn{obs})$. In addition to $n_\rmn{r}$, other statistics 
such as the length of the longest run or the symmetry between $N_+$ and $N_-$ may be checked.

For the fits to the full data sets whose parameters are listed in Tables~\ref{tab:conf1}
and~\ref{tab:conf2}, Tables~\ref{tab:run1} and~\ref{tab:run2} show the result of the
corresponding run tests over the full light curve and over the peak region only.
Reqiring an associated probabilitiy $P_\rmn{r} \geq 0.05$ only lets the single-lens
models for $d=0.1$ and $\sigma = 1~\%$ or $\sigma = 0.5~\%$ for both configurations survive,
while also the model for $d=0.15$ and $\sigma = 1~\%$ for configuation I nearly makes it to the
acceptable region if $P_\rmn{r} \geq 0.01$ has to be fulfilled.
If one looks at the difference between single and binary-lens light curves, one sees that
it becomes undetectable as soon as it is overshadowed by the statistical spread of the data
which appears to be the case for $d=0.1$ and $\sigma_0 = 1~\%$. In fact, the $\chi^2$ test is not
limited by this effect, but in principle allows to detect signals below the noise level by
means of a sufficiently large number of independent measurements.

The simulations show that single lens models for the discussed configurations involving a
K or M Bulge giant
look acceptable for $d=0.1$, unless the photometric
uncertainties are pushed significantly below the $1~\%$-level. The corresponding offset in
the limb-darkening coefficient of $\sim\,10~\%$ implies that the assumption of a single lens
is incompatible with the desire of a precision measurement on $\Gamma$. In contrast, the 
effect on the time-scale $t_\star$ is below $1\,\%$, so that potential lens binarity does not
have a significant effect on the measurement of the proper motion $\mu = \theta_\star/t_\star$,
given that the uncertainty in $\theta_\star$ is much larger.
One might want to argue that the binary lens system is an unlikely configuration. 
However, $d=0.1$ means a projected separation $a_\rmn{p} =  0.2~\mbox{AU}$ for
configuration I with two lens stars of mass $M/2 \sim 0.18~M_{\sun}$ and
$a_\rmn{p} =  0.3~\mbox{AU}$ for configuration II with two lens stars of mass
$M/2 \sim 0.35~M_{\sun}$ which should not a-priori be discarded.

\section{Final conclusions and summary}
\label{sec:conclusions}

Generally speaking, the potential binarity of the lens limits the accuracy of
a limb-darkening measurement of the observed source star for close-impact microlensing
events. The caustic near the center of the lens star is never exactly point-like, but always 
a small diamond with four cusps. If a point lens is
assumed, the obtained limb-darkening coefficient is systematically offset, while
the nature of the lens binarity may not be apparent. Although the inclusion of
a binary lens
in the modelling of the event will yield a proper limb-darkening 
measurement, the additional degrees of freedom still diminish the achievable 
precision. Unfortunately, such a computation is extremely demanding.
Simulations for typical event configurations involving K or M Bulge giants show that single-lens
models for binary lens events that involve an offset of  $\sim\,10~\%$ on the limb-darkening
coefficient $\Gamma$
look acceptable both from $\chi^2$ and from run tests, unless photometry significantly below the
$1~\%$-level is possible. The measurement uncertainty of the proper motion
$\mu = \theta_\star/t_\star$ however is dominated by the determination of the angular source radius
$\theta_\star$, involving stellar spectra, compared to which the accuracy limits on $t_\star$ caused
by lens binarity for close-impact microlensing events are negligible.

In contrast,
the measurement of limb darkening from fold-caustic passages does not
suffer from any of the problems encountered for close-impact events.
A binary lens is assumed a-priori, but the
measurement of limb darkening only depends on a smaller set of local properties
rather than on the complete binary-lens parameter space. The use of the local 
approximation of the light curve in the vicinity of the fold-caustic passage 
\citep[e.g.][]{PLANET:SMC,Do:Fold} makes
the computation quite inexpensive and easy. 
Moreover, fold-caustic exits can be well predicted. Once a caustic entry has been observed,
it is clear that a corresponding exit will occur, and the light curve on the
rise to the caustic exit peak 
allows to predict the time of the caustic exit with sufficient precision 
usually more than a day in advance.
A fair coverage of the caustic entry as well as the determination of the 
spectral type of the source star and an early measurement of the event time-scale $t_\rmn{E}$
even allows a rough guess on the passage duration and therefore a proper a-priori assessment
of the potential for measurements of the source brightness profile.
While the characteristic properties of a fold-caustic exit,
which provides a full scan of the source
from the leading to the trailing limb, can therefore be estimated in advance,
corresponding predictions
for close-impact events are only possible after the leading limb has alreaded passed the caustic,
leaving only the second half of the caustic passage involving the trailing limb.
From fold-caustic passages,
a linear limb-darkening coefficient can be obtained routinely and easily with a precision of less
than $5~\%$
\citep[e.g.][]{Do:FoldLD}.
Hence, events involving fold-caustic
passages turn out to be the clear favourite over close-impact events
for measuring limb-darkening coefficients
apart from those where the source transits a cusp. The latter provide an opportunity to determine
further limb-darkening coefficients beyond a linear law, which is quite challenging and usually
impossible for fold-passage events \citep{Do:SecondLD}.

However, events with fold-caustic passages make the worse case for proper motion measurements.
Fits to the light curve in the vicinity of the caustic passage only yield 
$t_\star^\perp = t_\star/(\sin \phi)$ \citep[e.g.][]{EROS:SMC,PLANET:SMC,Do:Fold},
where $\phi$ is the caustic-crossing angle,
which needs to be determined
from a model involving the full set of binary lens parameters. 
While a precise measurement of  $t_\star^\perp$ is possible, the uncertainty in
$t_\star$ is severely limited by degeneracies and appararent ambiguities for the
caustic-crossing angle $\phi$ \citep{Do99:CR,PLANET:sol,joint}, which can have a comparable
or even larger influence on the determination of the
proper motion $\mu$ than uncertainties in the angular stellar radius $\theta_\star$.

\bibliographystyle{mn2e}
\bibliography{binvsld}

\end{document}